\documentclass[aps,pre,twocolumn,showpacs,groupedaddress]{revtex4}
\usepackage{graphicx}

\begin{document}

\title{Validity of the scaling functional approach for polymer interfaces\\ as a
variational theory}

\author{Manoel
Manghi}\email{manghi@theorie.physik.uni-muenchen.de}\altaffiliation{Sektion
Physik, Ludwig Maximilian University, Theresienstr. 37, 80333 Munich,
Germany.}\author{Miguel Aubouy}\email{maubouy@cea.fr}

\affiliation{\mbox{Groupe Th\'eorie,
SI3M, UMR 5819 (CEA-CNRS-Univ. J. Fourier),}\\ D\'epartement de Recherche
Fondamentale sur la Mati\`ere Condens\'ee,\\ CEA-Grenoble, 38054 Grenoble cedex
9, France.}

\date{\today}

\begin{abstract}
We discuss the soundness of the scaling functional (SF) approach proposed by
Aubouy Guiselin and Rapha\"{e}l (Macromolecules \textbf{29}, 7261 (1996)) to
describe polymeric interfaces. In particular, we demonstrate that this
approach is a variational theory. We emphasis the role of SF theory as an
important link between ground-state theories suitable to describe adsorbed
layers, and "classical" theories for polymer brushes.
\end{abstract}


\maketitle

\section{Introduction}

Polymer interfaces are layers made of polymeric chains in direct contact
with a boundary which may be a solid/liquid, liquid/liquid interface or a
more complex surface such as a membrane. Because they have applications in
such diverse fields as colloid stabilization, coating, tribology, galenic,
they have been the subject of active research since the 80's both from a
fundamental and applied point of view. At present, there are two well
established self-consistent-field (SCF) theories to describe polymer layers.
They both start from the partition function of an ensemble of chains in contact
with the interface treated in mean-field, but they soon proceed in a marked
different way. Eventually, they become very different type of theories,
depending on whether the chains are reversibly adsorbed, and there is an
adsorbed state which dominates the solution of the Schr\"{o}dinger equation
associated (ground state dominance (GSD)
theories~\cite{holbook,Semenov-Joanny}), or they are end-tethered to a repulsive
surface (so-called ''brushes''), and the path integral is dominated by the
classical solution (classical theories~\cite{Semenov,MWC,Zhulina}).

Because the two types of theories are very different in spirit, there is a
conceptual gap for intermediate cases. In other words, there is no
mean-field theory available to describe both adsorption and grafting of
polymers within the same formalism. Such case arises, e.g., when chains are
grafted onto an attractive surface. In principle, at least, one should be
able to go in a continuous way from adsorbed-like to brush-like layers by
tuning the amount of chains per unit surface.

A tentative to bridge such gap was proposed in a series of paper where the
so-called Scaling Functional (SF) approach is developed~\cite{AGR,se}. This is
an approach where the layer of monodisperse adsorbed chains ($N$ monomers of
size $a$) is considered as a thermodynamic ensemble of interacting loops and
tails. These loops are polydisperse in size, and the main tool is the "loop
size profile", $S$, such that
\begin{equation}S(n)=S_{0}\int_{n}^{N}P(u)du,
\label{defS}
\end{equation}
where $P$ is the statistical distribution of loop sizes in monomer units, and
$S_{0}$ is the total number (per $cm^{2}$) of loops. The free-energy (per
$cm^2$) of the layer of chains is written as:
\begin{eqnarray}
\mathcal{F}\{S\} &\cong& \frac{k_BT}{a^2} \int_0^N
\left\{k[a^2S(n)]^{\beta}\right. \nonumber \\ &+& \left.[-a^2S'(n)] \ln
\left[-\frac{S'(n)}{S_0}\right]\right\} dn, \label{fenergie}
\end{eqnarray}
where $k\cong 1$ is a constant, $k_BT$ is the thermal energy and
$S'(n)=dS/dn$. The first term in rhs of
Eq.~(\ref{fenergie}) accounts for loop interactions (which depend on solvent
conditions through the value of the exponent $\beta$, see Table~\ref{table}).
The second term in the rhs of Eq.~(\ref{fenergie}) is the usual entropy
associated with a set of polydisperse objects. Similarly, the extension of the
layer is computed following
\begin{equation}
L\{S\} \cong a\int_0^N[a^2S(n)]^{\alpha}dn ,
\end{equation}
where the exponent $\alpha$ is given in Table~\ref{table}. In the SF
approach, the layer of chains is actually described as a polydisperse polymer
brush (the role of the chains being played here by the "pseudo-loops", i.e. half
loops) \textit{plus} an entropic term which stems for the fact that the size
distribution is not fixed by any external operator, but the system of loops is
in thermodynamic equilibrium.

\begin{table}
\begin{tabular}{|c|c|c|c|c|}
\hline type of solvent & good & $\Theta$ & melt & "mean-field" \\
\hline $\alpha$ & 1/3 & 1/2 & 1 & 1/3 \\
\hline $\beta$ & 11/6 & 2 & 3 & 5/3 \\
\hline
\end{tabular}
\caption{\label{table} Values of the scaling exponents for the layer
thickness and the free energy.}
\end{table}

If we impose monodisperse pseudo-loops ($P(u)=\delta (u-N)$) and $S_0=\sigma$,
the grafting density, we immediately recover the standard results for polymer
brushes. In good solvent conditions, these are: the extension $L\cong
aN(a^2\sigma)^{1/3}$, the free energy $\mathcal{F}\cong k_BTN(a^2\sigma)^{11/6}$
and the volume fraction of monomers $\Phi \cong (a^2\sigma)^{2/3}$. On the other
hand, if we let the polydispersity free to minimize the thermodynamical
potential (with $S_0=a^{-2}$ to account for attraction), we recover the results
found for reversibly adsorbed chains. In good solvent conditions, we find that
the volume fraction of monomer scales as $\Phi(z)\cong (a/z)^{4/3}$, and the
extension as $L\cong aN^{3/5}$.

Such idea proved to be successful in describing many different kinds of
polymer layers (grafted, reversibly adsorbed~\cite{AGR}, irreversibly
adsorbed~\cite{Guiselin}), whatever the solvent quality (good solvent,
$\Theta$-solvent and melt, i.e. no solvent). The approach was further expanded
to the cases of convex interfaces~\cite{AGRspheres}.

The success of this phenomenological approach lead us to address the
status of Eq.~(\ref{fenergie}). The SF approach is so far an elegant model but
not a theory because Eq.~(\ref{fenergie}) is not deduced from first principles,
and the set of approximations involved is not explicited. Recently, the SF
approach was applied to the issue of surface tension of polymeric
liquids~\cite{ManoPRL,ManoMacromol,ManoColloid}. Here again, the SF approach
proved to be successful in reproducing the experimental features in great
detail. However, because the results presented in Ref.~\cite{ManoPRL} are
different from the results of the self-consistent field theory on the same
issue, it seems important to clarify the soundness of the SF approach. This
question is addressed here in some detail.

The SF approach raises two questions essentially: \textit{a)}~is it sound ?
\textit{b)} is it valid ? The first question addresses the status of the SF
approach, the second has to do with the validity of the results that we will
find by using it. Obviously, these two issues are linked. Because "sound"
is sometimes used for "crude" or "inaccurate", it is useful to carefully
explain what we mean by "sound" and "valid" before we start arguing.

As it stands, the SF approach is a phenomenological description. This is
useful on issues where we do not have any theory available. On the other
hand, suppose we are in a position to compare a phenomenological approach
to a theory on the same issue. The theory will always prevail. If the two
results are in agreement, this is fine, but then the phenomenology is a
trick to understand qualitatively the issue, and essentially does not bring
new features. If, on the contrary, the two results are different, there is
always the suspicion that the phenomenological approach is a good idea
extrapolated to an issue where this idea is too simple, and therefore, the
result is wrong. We simply say "the approach is not sound". Accuracy then is
less relevant.

The debate is quite different when we have to compare two theories on the
same issue. If somehow we were able to deduce the SF approach from first
principles, and therefore prove that this is a theory, then the question of
soundness is resolved. Of course this would be done within approximations,
and the theory may be crude or inaccurate to treat the issue, but it is sound.
Then the debate over accuracy is essential to evaluate the results.

We see that the status of the SF approach is the first question to be
addressed, and depending on the answer, the debate over validity will be
different. In Section~\ref{status}, we deduce the effective free energy,
Eq.~(\ref{fenergie}), from first principles. In doing so, we demonstrate that
the SF approach is indeed a variational theory for polymer layers. Then we are
lead to ask the second question: is it valid~? Such task involves comparing the
results found with SF theory to SCF theories both at a formal level, and at the
level of the results. In Section~\ref{validity}, we address this question.

\section{Status \label{status}}

\subsection{Variational free energy}

We consider a set of $N_C$ monodisperse, linear, neutral chains in contact
with a solid plane (area $\Sigma $). We assume that the layer is uniform in
the directions parallel to the surface. Our starting point is the partition
function, $\mathcal{Z}$, of the chains, each characterized by the path $z_i(n)$,
where $z_i$ is the position normal to the surface and $n$ is the curvilinear
index ($1\leq i\leq N_C$):
\begin{equation}
\mathcal {Z} =\prod_i^{N_C} \int_0^{\infty} dz_i(0)
\int_0^{\infty} dz_i(N)\int \mathcal{D}\{z_i\}\exp \left[ - \frac{{\mathcal
H}}{k_BT}\right]
\end{equation}
where the effective Hamiltonian, $\mathcal{H}$, is the sum of an elastic
(entropic) contribution,
\begin{equation}
\mathcal{H}_{\mathrm{el}}=\frac32\frac{k_BT}{a^2}\sum_i^{N_C}\int_0^N
\left(\frac{dz_i}{dn}\right)^2 dn
\end{equation}
and an excluded-volume (two-body) interaction with parameter $v$,
\begin{equation}
\mathcal{H}_{\mathrm{ex}}=\frac{vk_{B}T}{2\Sigma}\sum_{i,j}^{N_C}
\int_0^N\int_0^N\delta \left(z_i(n)-z_j(n')\right) dndn'
\end{equation}
where $\delta$ is the Dirac distribution. We limit ourselves to two-body
interactions and thus neglect interactions of further order. This is not
valid for a $\Theta$-solvent (where $v=0$), but as we shall see in
Section~\ref{solvent}, the correct free energy for this type of solvent is
easily introduced afterwards. The volume fraction at distance $z$ writes
\begin{equation}
\phi(z)=\frac{1}{\Sigma}\sum_{i=1}^{N_C}\int_0^N\delta(z-z_i(n))dn.
\label{philoc}
\end{equation}

\begin{figure}
\includegraphics{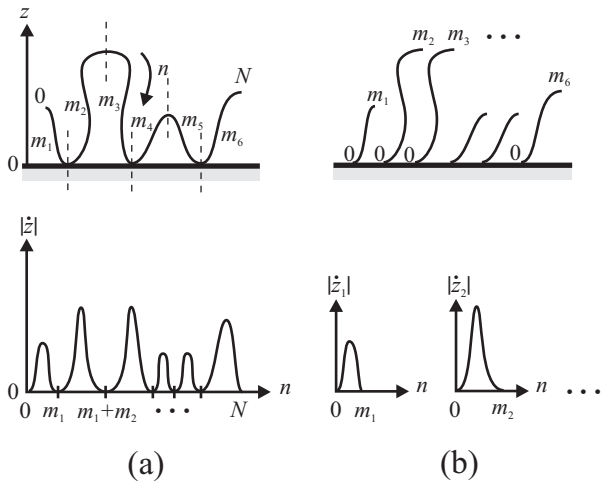}
\caption{\label{figure} As far as the Hamiltonian is concerned, each
chain in contact with the boundary (a) with associated path
$z(n)$ ($1<n<N$), is formally equivalent to the set of pseudo-loops (b) obtained
by cutting the loops into two equal pieces with associated paths
$\{z_{\alpha}(n),1<n<m_{\alpha}\}$.}
\end{figure}

Regardless of the particular microscopic situation that is realized, we can
always decompose the chain into loops and tails, and rewrite $\mathcal{H}$
accordingly. This amounts to cut the integrals into smaller pieces, by
identifying the monomers either in contact with the surface or at the top of the
loops and that we note $n_{i,\alpha}$. Each piece corresponds to the complete
path of a loop or a tail. The "cutting" scheme is described in
Fig.~\ref{figure}. We implicitly assume that the loops are symmetric, which comes
from the translation invariance parallel to the solid surface. Hence
mathematically, these identified monomers have a "null velocity":
$dz_i/dn|_{n_{i,\alpha}}=0$ for each $i$ and $\alpha$. The chain $i$ is then cut
in $N_i$ pieces of size $m_{i,\alpha}=n_{i,\alpha}-n_{i,\alpha-1}$
where $1\leq\alpha\leq N_i$ with $\sum_{\alpha }m_{i,\alpha}=N$. For tails, we
consider the full path from the extreme monomer to the first monomer in direct
contact with the surface, as expected. The loop are cut into two
pieces of equal length, which we shall call pseudo-loops. Clearly, as far as
mathematics is concerned, tails and pseudo-loops are similar objects : these are
chain segments starting at the surface and ending
somewhere in the solution  with no velocity at these extreme monomers (cf.
Fig.~\ref{figure}). For that reason, we shall not distinguish between tails and
pseudo-loops in the rest of the letter, and refer to both of them as
"pseudo-loops". As is obvious, such decomposition \textit{a)} is always
possible, \textit{b)} is unambiguous, \textit{c)} let the partition function
$\mathcal{Z}$ identical without any approximation, provided that we supplement
the cutting procedure by the constraint (later referred to as $\mathcal{C}$),
that the free extremities of the chain \textit{segments} originating from the
same loop should be at the same height $z$. Then $\left(m_{i,\alpha},
\left\{z_{i,\alpha}\right\}\right)_{\alpha=1,N_i}$ designates the set of sizes
and paths of pseudo-loops for chain $i$, and we rewrite the Hamiltonian:
\begin{widetext}
\begin{equation} \mathcal{H} =
\frac{3}{2}\frac{k_{B}T}{a^{2}}\sum_{i=1}^{N_{C}}\sum_{\alpha
=1}^{N_{i}}\int_{0}^{m_{i,\alpha }}\left( \frac{dz_{i,\alpha }}{dn}\right)
^{2}dn + \frac{vk_{B}T}{2\Sigma
}\sum_{i,j=1}^{N_{C}}\sum_{\alpha ,\beta =1}^{N_{i}}\int_{0}^{m_{i,\alpha
}}\int_{0}^{m_{j,\beta }}\delta \left( z_{i,\alpha }(n)-z_{j,\beta }(n^{\prime
})\right) dndn^{\prime } .
\label{Hdécomposé2}
\end{equation}
\end{widetext}

Computing exactly the partition function of the system with the Hamiltonian
Eq.~(\ref{Hdécomposé2}) is clearly out of reach. Rather, we implement the
variational principle which necessitates two steps~\cite{variationnel}.
First, we need to choose a trial probability such that $\mathcal{P}_{T}$
is a good approximation of the actual probability, $\mathcal{P}=
\mathcal{Z}^{-1} \exp [-\mathcal{H}/k_{B}T]$, but nevertheless allows for
analytical calculations. Second, we approximate the exact free-energy,
$\mathcal{F}$, of the system by the extremum of the functional
$\mathcal{F}_{\mathrm{var}}\{\mathcal{P}_{T}\} = \left\langle
\mathcal{H}\right\rangle_{\mathcal{P}_{T}}+ k_BT\left\langle
\ln\mathcal{P}_{T}\right\rangle_{\mathcal{ P}_{T}}$. Of these two steps,
the second one is the simplest because it is purely a matter of calculation.
Only the first one is significant as regard to the physics, since the success of
the variational theory lies in finding an appropriate trial function. The guess
$\mathcal{P}_{T}$ is a functional form with free unspecified parameters. By
minimizing $\mathcal{F}_{\mathrm{var}}\{\mathcal{P}_{T}\}$ with respect to these
parameters, we will obtain $\mathcal{P}_{T}$ with the chosen functional form
that best approximates $\mathcal{P}$. This is what ultimately controls
the difference between $\mathcal{F}$ and the approximation $\mathcal{F}_{\mathrm{var}}$.
Note that the choice of the ensemble of functions over which we shall perform
the minimization is arbitrary. It is a guess, not an approximation which could
be somehow quantified \textit{a priori}.

Our guess for $\mathcal{P}_{T}$ is:
\begin{equation}
\mathcal{P}_{T}\left(\left\{(m_{i,\alpha},
\{z_{i,\alpha}\})_{\alpha=1,N_i}\right\}_{i=1,N_C}\right)=
\prod_{i=1}^{N_C}\prod_{\alpha=1}^{N_i}P(m_{i,\alpha},\{z_{i,\alpha}\}),
\label{defP}
\end{equation}
where $\int_0^N P(m_{i,\alpha},\{z_{i,\alpha}\})dm=1$. Equation~(\ref{defP}) is
a mean-field type of approximation for the pseudo-loops since their
probability distributions are decorellated (hypothesis~\textit{A}). Furthermore,
we assume that \textit{the path $\{z_{i,\alpha}\}$ is the same for all the
pseudo-loops and is noted $\{z\}$} (hypothesis~\textit{B}). Because
$P(m_{i,\alpha},\{z_{i,\alpha}\})$ does not depend on the particular pseudo-loop
that is considered, we can drop the indexes and write $P(m,\{z\})$. Hence the
probability distribution reads $\mathcal{P}_{T}=P(m,\{z\})^{B}$ where
$B=\sum_{i=1}^{N_C}N_i$ is the number of pseudo-loops at the interface. The
crucial point is that $\mathcal{P}_{T}$ does no more depend on the complete set
of sizes and path, $\left\{ m_{i,\alpha },\left\{ z_{i,\alpha }\right\} \right\}
$, but only on \textit{a)} the size of the pseudo-loop, $m$, and on \textit{b)}
the path $z$, chosen to be the same for all pseudo-loops. Importantly, the
constraint $\mathcal{C}$ is automatically fulfilled with our approximation
since two pseudo-loops originating from the same loop have the same size, $m$,
and thus terminate at the same height $z(m)$. Then, the system is described by
two functions: $P(m)$, the probability that we have a pseudo-loop of size $m$,
and $z(n)$, the path of the chain segments. Hence, the trial free energy is
obtained by minimizing $\mathcal{F}_{\mathrm{var}}$ with respect to changes in
$z$ and $P$ (later, we will find it more convenient to work with $S$, rather
than $P$).

With~(\ref{defP}), we find
\begin{equation}
\left\langle\mathcal{H}_{\mathrm{ex}}\right\rangle_{\mathcal{P}_{T}} =
\frac12 \frac{v\Sigma}{a^6} k_BT \int
dz\Phi^2(z)
\label{Hex}
\end{equation}
where
\begin{eqnarray}
\Phi(z) &\equiv& \left\langle\phi(z)\right\rangle_{\mathcal{P}_{T}} \nonumber
\\ &=& a^3S_0\int_0^N dmP(m)\int_0^m\delta(z-z(n))dn
\label{phim}
\end{eqnarray}
and
\begin{equation}
\left\langle \mathcal{H}_{\mathrm{el}}\right\rangle
_{\mathcal{P}_{T}}=\frac32 \frac{k_BT}{a^2}\Sigma S_0
\int_0^Ndm P(m)\int_0^m\dot{z}^2(n)dn
\label{Hel}
\end{equation}
where $B=\Sigma S_0$ (hence $S_0$ is the "grafting density" of pseudo-loops),
and $\dot{z}=dz/dn$. Similarly, the entropic part of $\mathcal{F}_{\mathrm{var}}$ is
found to be
\begin{equation}
k_BT\left\langle \ln \mathcal{P}_{T}\right\rangle_{\mathcal{P}_{T}} = k_BT\Sigma
S_0\int_0^NdmP(m)\ln P(m).
\end{equation}
Combining all these results and integrating by parts and using Eq.~(\ref{defS}),
we find
\begin{eqnarray}
\frac{{\mathcal F}_{\mathrm{var}}(\{S\},\{\dot{z}\})}{k_BT\Sigma} &=& \int_0^N
\left\{\frac{3}{2a^2}\dot{z}^2(n)S(n) +
\frac{v}{2}\frac{S^2(n)}{\dot{z}(n)} \right. \nonumber\\
&-& \left. S'(n)\ln\left(-\frac{S'(n)}{S_0}\right)
\right\}dn .
\label{centralresult}
\end{eqnarray}
Note that $\Phi(z)=S(n(z))/\dot{z}$. Equation~(\ref{centralresult}) is the
central result of this letter which we now discuss. To get the best
approximation, we minimize Eq.~(\ref{centralresult}) with respect to $\dot{z}$
which yields $\dot{z}=\left(va^{2}/6\right)^{1/3} S^{1/3}$, and when this
result is introduced back into Eq.~(\ref{centralresult}), we find
Eq.~(\ref{fenergie}) with $\beta =5/3$ and
$k=\frac{3.6^{1/3}}{4}\left(v/a^3\right)^{2/3}$. We thus find the mean-field
version of our effective free energy Eq.~(\ref{fenergie}) with a numerical
coefficient, $k$, of order 1.

The formal derivation presented here brings an interesting remark. In the
early developpements of the SF theory, the entropic part in Eq.~(\ref{fenergie})
was introduced (and interpreted) as a contribution arising from combinatorial
arrangements of pseudo-loops at the surface: the presence of the interface
breaks down the symmetry of the solution and these monomers in contact with the
surface become \textit{distinguishable}. We see that the entropic term in
Eq.~(\ref{centralresult}) is formally that contribution arising from the entropy
of the trial probability.

\subsection{Generalization to other solvent conditions \label{solvent}}

The generalization to other solvent conditions, i.e. good solvent,
$\Theta$-solvent and melt, has been done in other references~\cite{AGR,se} and
deserves some comments.

In the case of a melt, the excluded volume interactions are screened at all
scales, and our mean-field approximation for pseudo-loops is automatically
verified. The probability distribution is then related to the Green function of
a chain by $P(m)\propto G(0,z(m);m)$ where $z(m)$ is self-consistently
determined \textit{via} the constraint $\dot{z}(n)=S_0\int_n^NP(m)dm$
($\phi(z)=1$ everywhere in an incompressible melt).

For a good solvent, the osmotic Eq.~(\ref{Hel}) and elastic term Eq.~(\ref{Hex})
are easily renormalized, following the des~Cloiseaux law~\cite{desCloiseauxlaw},
and using semi-dilute blobs~\cite{PGGbook}. However, the approximation which
consists in neglecting correlations between pseudo-loops is \textit{a priori}
not verified. Thus, the transformation of $k_BT\langle\ln{\mathcal
P}_T\rangle_{{\mathcal P}_T}$ in $k_BT\Sigma S_0\int_0^Ndm P(m)\ln P(m)$ is not
justified. However, correlations between monomers inside the same pseudo-loop
are taken into account through the blob renormalization.

Hence we have demonstrated that the SF approach is a variational
theory, and Eq.~(\ref{fenergie}) is sound.

\section{Validity \label{validity}}

Of course, that the SF theory is sound (in the sense that it is deduced
from first principles) does not guarantee at all that it is accurate, or
even simply valid to describe polymeric layers. This is because we have made
approximations whose range of validity remains to be examined.

\textit{A priori}, we could distinguish three different point of view to discuss
the issue of accuracy: \textit{a)} internal, \textit{b)} external and
\textit{c)} experimental.

\subsection{Internal estimate of accuracy}

Internal means that we are able to estimate the error that we have made in
approximating the initial Hamiltonian, and thus propose an internal
criterium of validity, very much like the Lifchitz criterium of validity for
mean-field theories. This requires that we define a relevant parameter which
would quantify the difference between the initial and the approximated
Hamiltonian, i.e. the two assumption that we made.

Concerning hypothesis~\textit{A}, we know that the mean-field
approximation for the loops is not valid in good solvent conditions. This
implies that the last term of Eq.~(\ref{centralresult}) is wrong. However, the
renormalization with semi-dilute blobs of the first two terms
takes into account the swelling of the pseudo-loops (hence
correlations between monomers) on scales smaller than the pseudo-loop sizes.
Thus for loops at least larger than one blob size, the excluded volume
interactions are screened and this loops are decorellated. Hence, the entropic
term of Eq.~(\ref{centralresult}) is justified for a large number of the
pseudo-loops and even if it is not fully satisfying, this is the best way we can
take into account these correlations unless we are lead to use renormalization
group theory, which has been done for one chain but not for many
chains~\cite{Eisenriegler}.

The hypothesis~\textit{B} is the crudest assumption in our theory. We assume
that all pseudo-loops have the same \textit{mean} path $z(n)$. It is easy to
show that for a melt, we find by minimization $z_{\mathrm{eq}}(n) \simeq
n^{1/2}$, which is the best variational approximation with our probability
distribution Eq.~(\ref{defP}). This result is quite similar to the Flory theorem
$R \simeq aN^{1/2}$ for the extension of a polymer chain in a melt. Of course
this result is valid for large $n$, since for a random walk, fluctuations
around this value is proportional to $n^{-1/2}$. This result may not be valid
for small loops. However, with variational theories, the estimate of this
error is impossible.

\subsection{External estimate of accuracy}

External means that we compare the SF theory with another theory. For polymeric
layers, the obvious candidate is SCF theories. \textit{A priori}, there are two
ways to do that: \textit{a)} a formal comparison, \textit{b)} a comparison of
the results that we obtain on a given issue. A formal comparison is simple
when the two theories have a common language. Unfortunately, this is not the
case for SCF theories and the SF theory. The former is deduced from the initial
Hamiltonian through a mean-field type of approximation for monomer-monomer
correlations which is then applied to the problem of polymer at interfaces,
whereas the latter proceeds in \textit{first} rewriting the Hamiltonian for
chains at interfaces and \textit{then} using a mean-field approximation for
pseudo-loops. Because of this different order for these two steps, we do not
know the way to formally compare SCF and SF theories. Then we are left with
comparing the results.

There are two issues where such comparison is possible: \textit{a)} brushes in
the infinite stretching limit, "mean-field" solvent conditions, and \textit{b)}
reversibly adsorbed layers, "mean-field" solvent conditions. This issues are
conceptually important because we know exactly the solution of the SCF theory in
the asymptotic limit $N\rightarrow\infty$.

\subsubsection{Brushes}

As shown by Netz and Schick~\cite{Netz} and Li and Witten~\cite{Li}, the
theory of polymer brushes proposed simultaneously by Milner, Witten, and Cates
(MWC) and Zhulina \textit{et al.} in Refs.~\cite{MWC,Zhulina}, which consists
in keeping the classic path in the partition function, can also be considered as
a variational approach. However, the trial probability is different, and the
layer is described by two functions: $g$, such that $g(z_0)dz_0$ is the
probability that the chain free extremity belongs to the interval
$[z_0,z_0+dz_0]$, and $e$, such that $e(z,z_0)=|dz/dn|$ is the extension at
position $z$ for a chain whose free-extremity is situated at $z_0$. Paths
(described by $e$) are chosen such as polymers are grafted at one end (with
grafting density $\sigma $), i.e. $\int_0^{\infty}\frac{dz_0}{e(z,z_0)}=N$
(which leads to the so-called equal time argument). The variational free energy
(per $cm^2$) is~\cite{Netz}:
\begin{eqnarray}
\frac{\mathcal{F}_{\mathrm{MWC}}}{k_BT} &=&
\frac{v}{2}\int_0^{\infty}\Phi^2(z)dz \nonumber \\ &+& \sigma
\int_0^{\infty}dz_0 g(z_0)\int_0^{z_0}\frac{3}{2a^2}e(z,z_0)dz  \nonumber \\ &+&
\sigma \int_0^{\infty}g(z_0)\ln [g(z_0)]dz_0, \label{NRJSchick}
\end{eqnarray}
with $\Phi(z)=\sigma \int_z^{\infty}dz_0 \frac{g(z_0)}{e(z,z_0)}$. Note
that in the context of brushes, the entropic contribution in
Eq.~(\ref{NRJSchick}), which is similar to that in Eq.~(\ref{centralresult}), is
the entropy of the chain end distribution~\cite{Netz,g1,g2}. Simple arguments
show that the first two terms in the rhs of Eq.~(\ref{NRJSchick}) scale as
$N(a^2\sigma)^{5/3}$, whereas $\int g(z_0)\ln [g(z_0)]dz_0\sim 1$. Hence,
in the strong stretching limit, $N(a^2\sigma)^{2/3}\gg 1$, the entropic
contribution to $\mathcal{F}_{\mathrm{MWC}}$ is
negligible~\cite{Netz}. However, this term is conceptually important and
has a physical signification since $e(z_0,z_0)$ is the tension sustained by the
free chain-ends. Hence, we see that Eq.~(\ref{centralresult}) and
Eq.~(\ref{NRJSchick}) are formally very close, but the choices for,
respectively, $z(n)$ and $e(z,z_0)$ are different.

To compare the SF theory with the MWC theory, we concentrate on monodisperse
brushes (hence the entropic contribution in Eq.~(\ref{centralresult})
disappears) in the strong stretching limit (hence, we neglect the entropic
contribution in Eq.~(\ref{NRJSchick})). We find in equilibrium
$\mathcal{F}_{\mathrm{MWC}}^*=0.892\,\mathcal{F}^*$. We see that the
extremum of $\mathcal{F}_{\mathrm{MWC}}^*$ is lower and according to the
variational criterium, the MWC theory is a better approximation of the exact
free energy. See~\cite{Li,milner} for a thorough discussion of this difference.
It is related to the different choices for the paths where the MWC choice
(i.e. the equal time argument) is less restrictive. The reason is that in the SF
theory for brushes, we impose an additional constraint: all chain free
extremities are situated in the outer edge of the layer, in a fashion similar to
the Flory approach (or the Alexander-de Gennes, which is similar in spirit but
introduces the correct scaling exponents). Formally, this amounts to impose a
delta type of function for $g$, a restriction motivated by our desire to keep
the SF theory tractable in a wider range of situations. Eventually, we find the
same results for $L$ and $\mathcal{F}^*$ at the scaling level,
although the description of the volume fraction profile is more accurate in the
MWC theory.

\subsubsection{Adsorbed layers}

Presumably, the case of reversibly adsorbed polymers is more significant for our
purpose since our variational approach is based on a "loop description", which
is justified for the homogeneous adsorption.

If we go to reversible adsorption, we have to turn our attention to GSD
theory. Although desirable, it is not so simple to compare the SF theory
with GSD theories. There are two reasons for this: \textit{a)} the GSD
theory uses the analogy between the partition function $\mathcal{Z}$ and the
Green propagator in quantum mechanics, which does not allow a description in
"polymer trajectories"; \textit{b)} in this theory, the free energy is expressed
in terms of the mean monomer concentration $\Phi(z)$, a quantity not simply
related to our probability density $P(m).$ Indeed, the partition function of a
chain having one end at $z$ and the other free, $\mathcal{Z}(N,z)$, in the SCF
theory, is the solution of the Schr\"{o}dinger equation : $\partial
\mathcal{Z}/\partial N=\frac{a^2}{6}\partial^2 \mathcal{Z}/\partial
z^2-U\mathcal{Z}$, where the external potential, $U$, is the sum of the
attractive potential due to the surface, $U_{\mathrm{surf}}$, and the
self-consistent potential, $U_{\mathrm{SCF}}$. For adsorbed chains, there is a
ground state of negative energy $-\varepsilon Nk_BT$ which dominates the
solution, and (in the limiting case where $\varepsilon N\gg 1$) the free energy
approximates to:
\begin{equation}
\mathcal{F}_{\mathrm{GSD}}=k_BT\int_0^{\infty}dz\left[\kappa(\Phi)\left(
\frac{d\Phi}{dz}\right)^2 + U(z)\Phi(z)\right] , \label{GSD}
\end{equation}
where $\kappa(\Phi)=a^2/(24\Phi)$. As shown by Lifshitz and des
Cloiseaux~\cite{Lifshitz,des Cloiseaux}, the square gradient term in
Eq.~(\ref{GSD}) has essentially an entropic origin~\cite{note}, whereas the
polymeric nature of the liquid can be neglected in the molecular field
$U_{\mathrm{SCF}}(z)$ (which is estimated for a monomeric liquid). Then we are
lead to think that the elastic and entropic part in Eq.~(\ref{centralresult})
are related to the square gradient term, but we are not able to rewrite the
former as the latter at the moment.

In the absence of any clue to formally compare Eqs.~(\ref{centralresult})
and~(\ref{GSD}), we shall compare their results for infinite chains and
mean-field potential, a limit where the GSD theory happens to be exact. If we
minimize the free energy Eq.~(\ref{centralresult}) with the boundary conditions
$S(0)=a^{-2}$, $S(N\rightarrow\infty)=0$, we find $a^2
S_{\mathrm{eq}}(n)=k'^{3/2}/(n+k')^{3/2}$ where $k'=(3/(2k))^{4/9}$ which yields
$\Phi(z)\sim z^{-2}$, essentially the solution found by minimizing
Eq.~(\ref{GSD}). Similarly, we find that $\mathcal{F}^*\cong k_BT/a^2$ as with
GSD theory. Hence, we find a very good agreement for infinite chains.

That the agreement should be better (in the sense that both the scaling and the
concentration profile are identical) for adsorption than for brushes
reflects the validity of our initial assumption that all pseudo-loops have
the same path. As explained in Ref.~\cite{MWCpoly}, for very polydisperse
layers, we expect a stratification of the locations of the free chain ends
above the surface. This is because the free ends of a long chain locates
further away from the surface than that of a short chain to take advantage
of a lower osmotic pressure (the concentration decreases away from the
interface). In the continuum limit, this argument suggests that every
pseudo-loops are similarly extended, and therefore validates our guess.

\subsection{Experimental estimate of accuracy}

To evaluate the accuracy of a variational theory, the ultimate and major
argument is to compare the value of the free energy at its minimum to
experiments. The good candidate is thus the surface tension of
polymeric liquids, $\gamma$. We have shown in
Refs.~\cite{ManoPRL,ManoMacromol,ManoColloid}, that the SF approach allows the
calculation of the variations of $\gamma(N)$ in very good agreement with
experimental data found in the literature. This is a good test for the theory
which has been done both for melts and semi-dilute solutions (in
good solvent).

It is important to note that the SCF theory in the GSD approximation leads to
a different result for the melt surface tension. The finite chain correction
in that case is proportional to $N^{-1}$. We found a larger correction in $\ln
N/N^{1/2}$. An explanation of this discrepancy is that the SCF description
relates the surface tension to the gradients in volume fraction which are
localized in a very thin layer of thickness $a$ (indeed this approach is not
valid for large gradients). We argue that this dependence comes from the
chains reorganization on a larger layer of thickness the radius of gyration of a
chain. In this layer, the volume fraction is constant. Thus it cannot be
described by the SCF approach whereas the SF approach uses different tools,
namely $z(n)$ and $S(n)$, which allows such a description. Hence, for adsorbed
layers from a melt and a semi-dilute solution, we see that these two approaches
are quite different.

\section{Concluding remarks}

This article aims at clarifying the debate concerning the "soundness" of
the Scaling Functional approach. In view of this, the demonstration that the
SF approach is a variational theory is certainly the essential and most
significant result of this article (Section~\ref{status}). But we think we have
made clear a certain number of points (Section~\ref{validity}). These are:
\begin{enumerate}
\item The SF approach is a variational theory and therefore has the same
epistemological status as SCF theories for brushes ("classical" solution) and
adsorbed chains (GSD). Of course, the approximations made are different and each
of these theories has a different range of validity.
\item Because the SF theory is a variational theory, we are not able to properly
quantify the approximations that are involved, and therefore we are unable to
define the range of validity of this theory.
\item There is no way that we know to formally compare the two theories because
the first step in approximating the initial Hamiltonian are different.
\item Each time a direct comparison with SCF theory is possible, we find
\textit{a)} always the same scaling results, and \textit{b)} sometimes the same
analytical result. Thus we conclude that the SCF theory does not provide any
argument against the SF theory.
\end{enumerate}
That the GSD approximation to the SCF theory is formally justified and
quantifiable in "mean-field" solvent conditions does not guarantee that the
result that we find is accurate for real solvent conditions and notably for the
melt case. A description only in terms of volume fraction (see Eq.~(\ref{GSD}))
comes also from a variational argument~\cite{des Cloiseaux} and has not been
quantitatively justified in real systems. In other words, the GSD in the limit
$N\rightarrow\infty$, is the exact solution of the SCF theory, but still an
approximate solution of the initial Hamiltonian.

The crucial point regarding our approximations (cutting loops into two
tails and describing all the pseudo-loops with the same path) is whether the
distinction between loops and tails is important enough to modify the
conclusions of a simple theory in which it is neglected. When we are in a
position to directly evaluate the consequences of these approximations, we find
that this distinction does not affect the scaling results. It is interesting to
note that the distinction between loops and tails has been done
self-consistently for the SCF theory but is put \textit{ad hoc} for other type
of solvents~\cite{Semenov-Joanny}. Therefore we conclude that there is no valid
argument to support that these approximations are not sound, provided that we
remain at a scaling level of description.

Finally, we assume in this approach that a large number of loops are formed at
the interface. This impose both a sharp interface and the presence of many
adsorbed chains. Therefore, this theory does not apply to single
chain adsorption and to systems such as interfaces between incompatible polymers
or diblock copolymers, for which other approaches based on the SCF theory
have been developed~\cite{Helfand,Leibler}.

As a conclusion, the SF theory proposes a compromise between a precise
description of the polymeric layer, and a wide ranging scaling type of
theory valid for arbitrary polymer layers, various solvent conditions and
various geometries. Since it does not require a comparable amount of mathematics
and has a wider range of applicability than both theories, it is very likely
that the SF theory will become an important piece of our understanding of
polymeric interfaces.

\begin{acknowledgments}
We are grateful to B. Fourcade for stimulating conversations. We also
benefitted from discussions with \mbox{J.-F.~Joanny,} A.N.~Semenov and 
R.R.~Netz.
\end{acknowledgments}

\end{document}